\documentclass[sigconf]{acmart}
\usepackage{algorithmic}
\usepackage{graphicx}
\usepackage{textcomp}
\usepackage{xcolor}
\usepackage{subfigure}
\usepackage{threeparttable}
\usepackage{url}

\usepackage{mdwtab}
\usepackage{booktabs}
\usepackage{array}
\usepackage{mdwmath}
\usepackage{eqparbox}
\usepackage{multirow, multicol}
\usepackage{makecell}
\usepackage{tabularx}
\usepackage{diagbox}
\usepackage{siunitx}
\usepackage{verbatim}
\usepackage{pifont} 
\usepackage{bbding}

\newcommand{\cmark}{\textcolor{green!80!black}{\ding{51}}}
\newcommand{\xmark}{\textcolor{purple}{\ding{55}}}

\newenvironment{packeditemize}{ \begin{list}{$\bullet$}{ \setlength{\labelwidth}{4pt} \setlength{\itemsep}{0pt} \setlength{\leftmargin}{\labelwidth} \addtolength{\leftmargin}{\labelsep} \setlength{\parindent}{0pt} \setlength{\listparindent}{\parindent} \setlength{\parsep}{0pt} \setlength{\topsep}{1pt}}}{\end{list}}

\usepackage{tikz}
\newcommand{\covcircle}[1]{%
\begin{tikzpicture}[baseline=-0.5ex]
  \filldraw[gray!20] (0,0) circle (0.9ex);
  \ifnum#1=25
  \fill[black] (0,0) -- (90:0.9ex) arc (90:180:0.9ex) -- cycle;
  \fi \ifnum#1=50
  \fill[black] (0,0) -- (0:0.9ex) arc (0:180:0.9ex) -- cycle;
  \fi \ifnum#1=75
  \fill[black] (0,0) -- (-90:0.9ex) arc (-90:180:0.9ex) -- cycle;
  \fi \ifnum#1=100
  \fill[black] (0,0) circle (0.9ex);
  \fi
\end{tikzpicture}%
}

\usetikzlibrary{patterns}
\newcommand{\whitebox}{\tikz{\draw[black,fill=white] (0,0) rectangle (1,0.2);}}
\newcommand{\graybox}{\tikz{\draw[black,pattern=north east lines,pattern color=black]
(0,0) rectangle (1,0.2);}}
\newcommand{\blackbox}{\tikz{\draw[black,fill=black] (0,0) rectangle (1,0.2);}}

\renewcommand{\footnotetextcopyrightpermission}[1]{} 
\begin{document}

  \title{SoK: Semantic Privacy in Large Language Models}

  \author{Baihe Ma$^{1}$, Yanna Jiang$^{1}$, Chen Li$^{1}$, Xuelei Qi$^{3}$, Xu
  Wang$^{1}$, Guangsheng Yu$^{1}$, \\
  Qin Wang$^{1}$, Caijun Sun$^{2}$, Ying He$^{1}$, Wei Ni$^{1}$, Ren Ping Liu$^{1}$}

  \affiliation{ \smallskip \textit{$^{1}$University of Technology Sydney, Australia} $|$ \textit{$^{2}$Zhejiang Lab, China} $|$ \textit{$^{3}$Northeastern University, China}}

  
  
  
  
  
  
  
  
  
  


  \begin{abstract}
    Large Language Models (LLMs) are distinguished by their advanced ability to internalize
    the meaning embedded in the training data. This deep understanding
    introduces a significant privacy risk beyond verbatim memorization, as models
    may reveal learned sensitive content through paraphrasing or inference, thus
    breaching \textbf{Semantic Privacy}.
    This Systematization of Knowledge (SoK) introduces a lifecycle-centric
    framework to analyze how semantic privacy risks emerge across input
    processing, pretraining, fine-tuning, and alignment stages of LLMs. We
    categorize key attack vectors and assess how current defenses, such as differential
    privacy, embedding encryption, edge computing, and unlearning, address these
    threats. Our analysis reveals critical gaps in semantic-level protection, especially
    against contextual inference and latent representation leakage. We conclude
    by outlining open challenges, including quantifying semantic leakage,
    protecting multimodal inputs, balancing de-identification with generation quality,
    and ensuring transparency in privacy enforcement. This work aims to inform future
    research on designing robust, semantically aware privacy-preserving
    techniques for LLMs.
  \end{abstract}


  \maketitle

  \section{Introduction}
  \label{Introduction}

  Large language models (LLMs) exhibit strong generalization capabilities across
  a wide range of tasks~\cite{nazi2024large}. However, their capacity to
  memorize training data introduces a distinct class of privacy risks. 
  Unlike conventional models that primarily capture distributional patterns, LLMs are capable of reproducing specific training instances, not only verbatim, but also through semantically equivalent rephrasings or stylistic reconstructions~\cite{farquhar2024detecting}.
  This phenomenon gives rise to what is increasingly referred to as semantic privacy leakage, wherein sensitive information can be indirectly revealed without requiring exact string reproduction.

  The probabilistic nature of LLM outputs further compounds the challenge. 
  A model may reveal memorized content only under particular prompting conditions, often in ways that evade string–matching–based detection~\cite{liu2025compromising}.
  For example, while a password containing a birthdate such as “2000-04-01” may not be reproduced explicitly, the model might generate: “She tends to choose combinations involving her birth year and a spring date.” 
  In such cases, the leakage occurs at the semantic level and can facilitate downstream inference or re-identification.
  To distinguish semantic privacy from general data privacy, we define semantic
  privacy as:

  \smallskip
  \textit{\textbf{Semantic privacy} aims to protect sensitive attributes that are
  not explicitly stated in the data but can be inferred from the data, often by
  leveraging contextual or external knowledge.
  }
  \smallskip

  While both semantic privacy and data privacy aim to prevent the exposure or misuse of sensitive information and can be enhanced by established privacy-preserving technologies like anonymization, encryption, and DP~\cite{das2025security}.
  However, as summarized in Table~\ref{tab:privacy-comparison}, data privacy is primarily
  concerned with protecting raw data from direct exposure. In contrast, semantic
  privacy focuses on safeguarding the sensitive inferences and relationships that
  can be derived from the data.

  \begin{table*}
    [t]
    \centering
    \caption{Key Differences Between Data Privacy and Semantic Privacy}
    \label{tab:privacy-comparison}
    \renewcommand{\arraystretch}{1.1}
    \resizebox{0.95\linewidth}{!}{%
    \begin{tabular}{c|l|l}
      \toprule \multicolumn{1}{c}{\textbf{Aspect}} & \multicolumn{1}{c}{\textbf{Data privacy}}                                        & \multicolumn{1}{c}{\textbf{Semantic privacy}}                                \\
      \midrule \textbf{Protection goal}            & Protects the data itself (e.g., PII)                                             & \makecell[l]{Protects relationships and inferences drawn from data}          \\
      \hline
      \textbf{Threat model}                        & Direct attacks on training data                                                  & \makecell[l]{Inference attacks based on model behavior and outputs}          \\
      \hline
      \textbf{Leakage channel}                     & Direct exposure of raw data                                                      & \makecell[l]{Indirect leakage via sensitive attribute inference}             \\
      \hline
      \textbf{Adversary knowledge}                 & \makecell[l]{Requires internal model knowledge \\ or access to training context} & \makecell[l]{Leverages external or world knowledge}                          \\
      \hline
      \textbf{System lifecycle}                    & Preprocessing and data handling                                                  & \makecell[l]{Covers full lifecycle: pretraining, fine-tuning, and inference} \\
      \bottomrule
    \end{tabular}%
    }
  \end{table*}

\begin{figure*}[!htpb]
    \centering
    \includegraphics[width=0.7\linewidth]{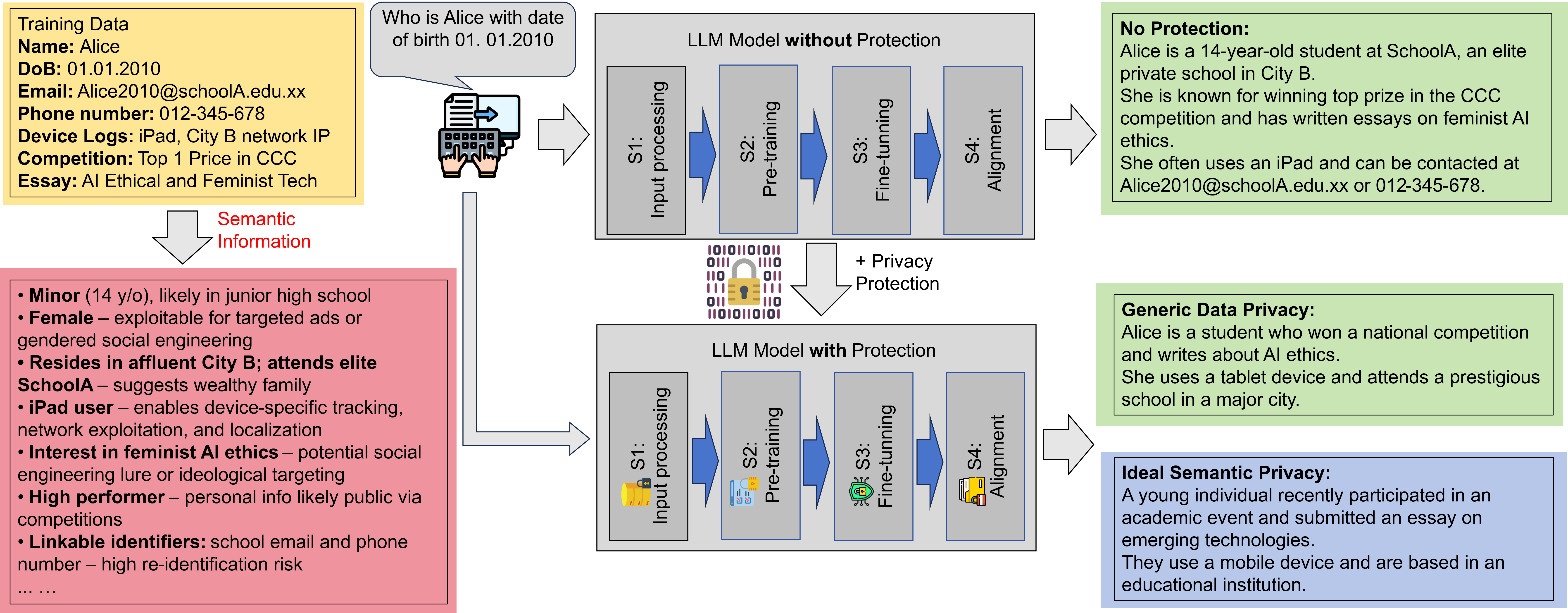}
    \caption{\textbf{Illustration of semantic privacy leakage across the LLM
    lifecycle.} The example demonstrates how rich semantic information can be inferred
    from natural inputs even after removing explicit identifiers (e.g., email
    and phone). Without protection, all sensitive details are directly exposed.
    Generic data privacy methods that strip surface identifiers still allow
    attackers to infer key latent attributes via semantic cues. In contrast,
    ideal semantic privacy aims to suppress inference pathways by disrupting the
    alignment between inputs and their latent semantics. This figure exemplifies
    our central insight: that semantic leakage arises from transformations
    across all stages of LLM processing, and surface-level anonymization is insufficient.}
    \label{fig:LLM-Semantic-Privacy}
  \end{figure*}

  When applying these concepts to LLMs, we define their scope as follows: \textbf{Data privacy} pertains to the protection of verbatim information contained within the training data. 
  \textbf{Semantic privacy} addresses a broader threat surface, covering sensitive semantic information that can be inferred at any stage of the LLM lifecycle—from the trained model's parameters and latent representations to its final generated outputs. 
  As illustrated in
  Figure~\ref{fig:LLM-Semantic-Privacy}, an LLM can generate text that paraphrases sensitive content from its training data, thereby leaking rich semantic information. 
  An ideal semantic privacy-preserving LLM would be capable of concealing this private semantic information.

  Existing works \cite{gan2024navigating, he2024emerged, neel2023privacy, yan2024protecting, das2024security, li2024llm} offered classifications of privacy threats and discussed techniques targeting data privacy in LLM. Yet, their treatment of semantic privacy remains limited.

  This paper aims to bridge this research gap by providing a comprehensive review
  of semantic privacy concerns and protection mechanisms in the context of LLMs,
  as follows:
\vspace{5pt}
  \begin{packeditemize}
    \item \textbf{RQ1:} What distinguishes semantic privacy from traditional data privacy in LLMs, and how do different stages of the LLM lifecycle contribute
    to semantic leakage?

    \item \textbf{RQ2:} How effective are current semantic privacy defenses (e.g.,
    DP, edge learning, encryption, unlearning), and what
    limitations do they face in mitigating inference from contextual, latent, or
    implicit semantics?

    \item \textbf{RQ3:} How can future research design quantifiable and adaptive
    defenses that preserve semantic privacy while maintaining LLM utility and interpretability?
  \end{packeditemize}

  \smallskip
  \noindent
  \textbf{Contributions.} We focus on semantic privacy risks in LLMs, addressing
  a critical gap in the current research, where traditional data privacy paradigms fall short. Our contributions are:
\vspace{5pt}
  \begin{packeditemize}
    \item We provide a formal definition of semantic privacy and establish its distinction
    from conventional data privacy, highlighting its unique threat landscape in
    LLMs.

    \item We propose a lifecycle-orientated analytical framework that
    systematically maps semantic privacy risks across the input, pretraining,
    fine-tuning, and alignment stages of LLMs. We also present a taxonomy of
    semantic privacy attacks and defences, synthesising fragmented literature and
    evaluating the limitations of existing protection techniques.

    \item We provide a structured gap analysis of previous surveys, highlighting
    the lack of a systematic treatment of semantic privacy, and position our work
    as the first semantically grounded systematization of this topic.
  \end{packeditemize}

  \smallskip
  \noindent
  \textbf{Key insights.} Building on this diagnostic perspective, we articulate
  a strategic roadmap for future research that moves beyond incremental
  improvements. Our key insights include:

  \begin{packeditemize}
    \item \textbf{Semantic privacy is threatened across the entire LLM lifecycle.}
    Attacks such as membership inference, attribute inference, model inversion, and
    backdoor triggering exploit vulnerabilities at every stage from input encoding
    to final alignment. These threats expose the limitations of surface-level anonymization and demand deeper, architecture-aware protections.

    \item \textbf{Existing defenses are fragmented and insufficiently aligned
    with semantic leakage pathways.} While techniques like DP,
    local processing, homomorphic encryption, and targeted forgetting address specific
    risks, none provide comprehensive coverage across representational layers. Strong
    semantic privacy requires stage-aware coordination and a balance between protection
    and performance.

    \item \textbf{Future solutions should be measurable, modality-aware, utility-preserving,
    and transparent:}
    \begin{itemize}
      \item[$\circ$] \textit{Quantification:} Traditional token-level perturbation methods inadequately capture the latent, context-driven nature
        of semantic leakage in LLMs. Future research should prioritize robust
        semantic privacy quantification by leveraging embedding-based similarity,
        entailment-aware scoring, and probabilistic re-identification metrics to
        assess risks stemming from meaning preservation, inference potential,
        and contextual traceability.

      \item[$\circ$] \textit{Multimodal modeling:} As LLMs increasingly operate
        across modalities (e.g., text, images, audio), privacy frameworks must account
        for cross-modal semantic entanglement, where sensitive information may
        be inferred through the interaction between modalities. Effective protection
        requires modeling modality interactions holistically, integrating
        attention-based suppression and modality-specific risk scoring to
        preemptively mitigate leakage pathways.

      \item[$\circ$] \textit{De-identification:} Semantic de-identification
        should move beyond entity-level anonymization toward adaptive, task-aware
        strategies—such as controlled semantic rewriting and privacy-preserving generation—that
        preserve fluency, coherence, and task fidelity while obfuscating latent identifiers.
        Personalized and explainable de-identification mechanisms will be
        essential for aligning technical efficacy with user expectations and
        normative accountability.
    \end{itemize}
  \end{packeditemize}

 \begin{table*}
    [!htpb]
    \centering
    \caption{Survey Comparison on Semantic Privacy Dimensions}
    \label{tab:multirow_survey}
    \begin{tabular}{c|cccc|cccc|cccccccc|c|c|c|c}
      \hline
      \multirow{2}{*}{\textbf{Reference}} & \multicolumn{4}{c|}{\textbf{Lifecycle}} & \multicolumn{4}{c|}{\textbf{Semantic threats}} & \multicolumn{8}{c|}{\textbf{Vulnerabilities}} & \multicolumn{4}{c}{\textbf{Other}} \\
      \cline{2-21}                        & S1                                      & S2                                             & S3                                            & S4                                & T1             & T2             & T3             & T4             & V1             & V2             & V3             & V4             & V5             & V6             & V7             & V8             & QF             & MM             & BA             & EX             \\
      \hline
      \cite{neel2023privacy}              & \covcircle{25}                          & \covcircle{25}                                 & \covcircle{25}                                & \xmark                            & \cmark         & \covcircle{25} & \covcircle{25} & \covcircle{25} & \covcircle{25} & \xmark         & \xmark         & \xmark         & \xmark         & \xmark         & \xmark         & \xmark         & \xmark         & \xmark         & \covcircle{25} & \covcircle{25} \\
      \cite{gan2024navigating}            & \covcircle{25}                          & \xmark                                         & \xmark                                        & \xmark                            & \covcircle{25} & \xmark         & \xmark         & \cmark         & \covcircle{25} & \xmark         & \covcircle{25} & \covcircle{25} & \xmark         & \xmark         & \covcircle{25} & \covcircle{25} & \covcircle{25} & \cmark         & \covcircle{25} & \covcircle{25} \\
      \cite{he2024emerged}                & \covcircle{50}                          & \covcircle{50}                                 & \covcircle{75}                                & \covcircle{75}                    & \covcircle{75} & \covcircle{75} & \covcircle{50} & \covcircle{50} & \covcircle{50} & \xmark         & \covcircle{50} & \covcircle{50} & \covcircle{25} & \covcircle{25} & \covcircle{25} & \covcircle{25} & \covcircle{25} & \covcircle{25} & \covcircle{25} & \covcircle{25} \\
      \cite{yan2024protecting}            & \covcircle{75}                          & \covcircle{50}                                 & \covcircle{50}                                & \covcircle{50}                    & \cmark         & \covcircle{25} & \covcircle{25} & \covcircle{25} & \covcircle{25} & \covcircle{25} & \covcircle{25} & \covcircle{25} & \covcircle{25} & \covcircle{25} & \covcircle{25} & \covcircle{25} & \covcircle{25} & \xmark         & \covcircle{25} & \covcircle{25} \\
      \cite{das2024security}              & \xmark                                  & \xmark                                         & \xmark                                        & \xmark                            & \covcircle{25} & \xmark         & \xmark         & \xmark         & \xmark         & \xmark         & \xmark         & \xmark         & \xmark         & \xmark         & \xmark         & \xmark         & \xmark         & \xmark         & \covcircle{25} & \xmark         \\
      \hline
      \textbf{Ours}                       & \cmark                                  & \cmark                                         & \cmark                                        & \cmark                            & \cmark         & \cmark         & \cmark         & \cmark         & \cmark         & \cmark         & \cmark         & \cmark         & \cmark         & \cmark         & \cmark         & \cmark         & \cmark         & \cmark         & \cmark         & \cmark         \\
      \hline
    \end{tabular}
    \begin{tablenotes}
      \item[] \textbf{Lifecycle:} S1 – Input processing; S2 – Pre-training; S3 –
      Fine-tuning; S4 – Alignment. \item[] \textbf{Semantic Threats:} T1 – Membership
      Inference; T2 – Attribute Inference; T3 – Model Inversion; T4 – Backdoor
      Attack. \item[] \textbf{Vulnerabilities:} V1 – Embedding; V2 – Tokenization;
      V3 – Attention; V4 – FFN; V5 – Normalization;

      \quad \quad \quad \quad \quad \quad \quad V6 – Output Heads; V7 - RLHF Modules;
      V8 - Generation Layer. \item[] \textbf{Others:} QF - Semantic
      Quantification; MM - Multimodal Risk; BA - Privacy–Utility Balance; EX - Explainability.
      \item[] \xmark: Not addressed; \covcircle{25}, \covcircle{50}, \covcircle{75}
      – Partial coverage; \cmark: Explicitly and systematically discussed.
    \end{tablenotes}
  \end{table*}
  \section{Existing Surveys and Gap Analysis}
  \label{Related work}

  This section surveys recent literature on LLM privacy, with a particular focus
  on the extent to which semantic privacy is addressed.

  Existing surveys (Table~\ref{tab:multirow_survey}) (e.g.,
  \cite{gan2024navigating}\cite{he2024emerged}\cite{neel2023privacy}) provided overviews
  of privacy threats, but treat semantic risks only implicitly or fragmentarily.
  For instance, Gan et al.~\cite{gan2024navigating} presented a lifecycle-based threat
  taxonomy yet overlooked how semantically rich interactions (e.g., context
  chaining and memory in multi-turn dialogues) facilitate indirect inference.
  Similarly, He et al.~\cite{he2024emerged} focused on agent-based
  vulnerabilities without examining how semantic reasoning enables privacy
  violations through natural language cues. While Neel~\cite{neel2023privacy} introduced
  foundational threat categories like membership inference, it does not address
  the role of contextual semantics in privacy leakage.

  Other studies, including Yan et al.~\cite{yan2024protecting} and Das et al.~\cite{das2024security},
  offered broader taxonomies and mitigation strategies, yet primarily frame privacy
  through data-centric lenses such as DP and federated learning.
  These techniques are effective in limiting raw data exposure, but are insufficient
  against semantic leakage, where sensitive information is reconstructed through
  language patterns, context, or reasoning. Even when inference-related threats are
  acknowledged, as in Li et al.~\cite{li2024llm}, mechanisms that operate at the
  semantic level (e.g., privacy-aware generation control or semantic obfuscation)
  remain underexplored. The inferential capacity of LLMs to deduce user intent,
  latent attributes, or background knowledge from semantic patterns remains an open
  threat vector.

  These limitations underscore several critical research gaps. First, existing privacy
  mechanisms failed to account for semantic-level correlations, which standard
  DP methods, premised on independent data records, cannot capture~\cite{neel2023privacy,yan2024protecting}.
  Second, runtime safeguards for semantic leakage, such as adaptive decoding or context-sensitive
  risk modulation, are largely absent~\cite{li2024llm}. Our work addresses these
  gaps by offering a clear definition of semantic privacy, a lifecycle-aligned
  analytical framework, and a semantic-centric evaluation of attacks and defenses,
  contributing a forward-looking agenda for privacy protection in multimodal and
  generative settings.

  \section{Semantic Privacy Risks}
  \label{Location privacy requirements}

\begin{figure*}[!htpb]
    \centering
    \includegraphics[width=0.7\linewidth]{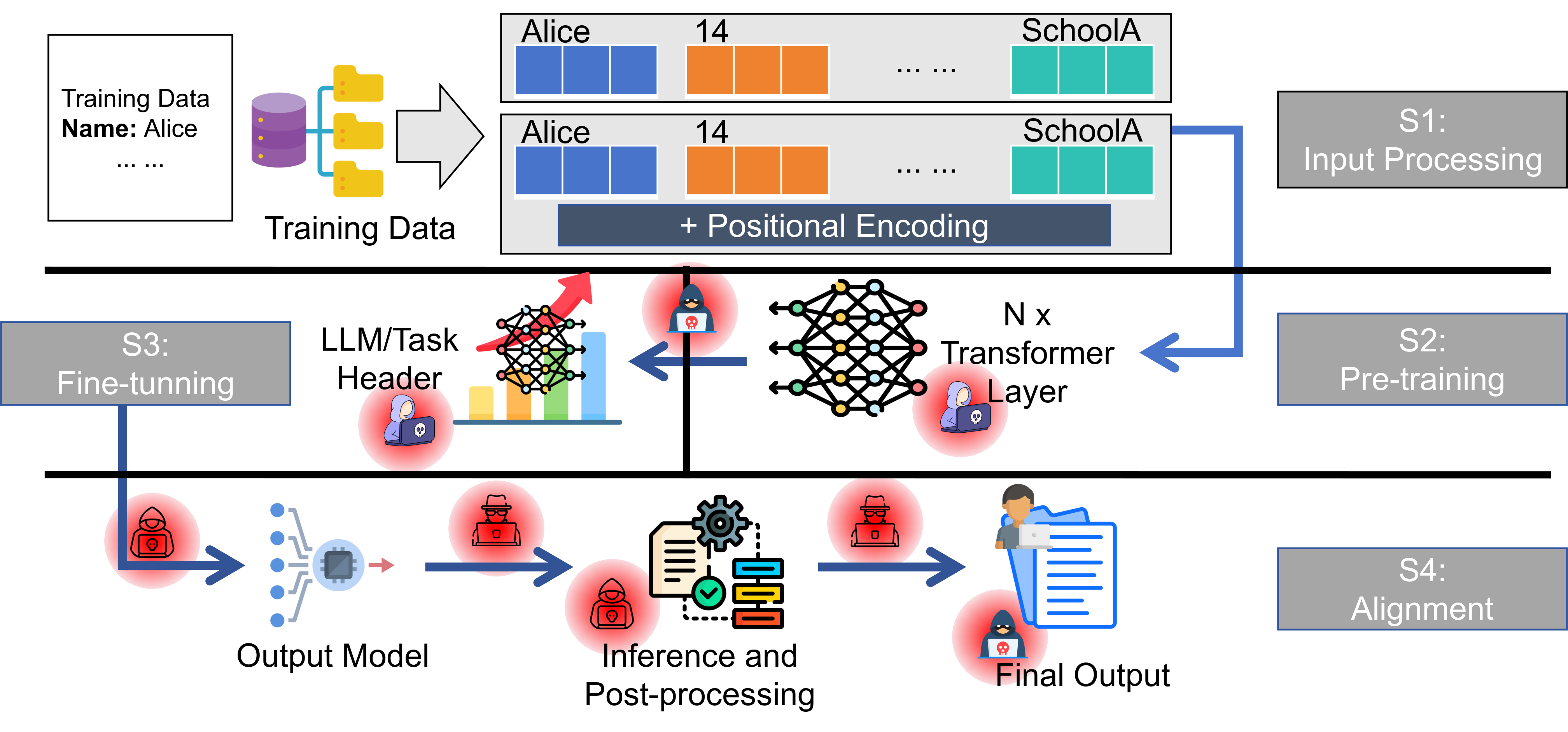}
    \caption{\textbf{Illustration of semantic information flow and corresponding
    attack surfaces throughout the LLM lifecycle.} As the training data is transformed
    through S1, S2, S3, and S4, semantic representations, such as identity, affiliation,
    or intent, are preserved. These internal representations become potential
    attack vectors for adversaries at each stage. Model inversion attacks
    \includegraphics[height=1em]{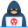}
    exploit the transformer layers in S2 to reconstruct sensitive training content;
    backdoor attacks \includegraphics[height=1em]{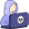}
    manipulate hidden triggers within the model during or after fine-tuning (S3);
    attribute inference attacks \includegraphics[height=1em]{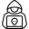}
    target semantic traits encoded in task-specific heads or outputs; and
    membership inference attacks \includegraphics[height=1em]{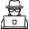}
    leverage output behaviors and post-processing artifacts to detect training
    data presence. }
    \label{fig:LLMstructure&Attack}
\end{figure*}

  LLMs rely on a multi-stage processing pipeline that transforms natural
  language inputs into semantically meaningful outputs. At each stage, the model
  extracts, preserves, and refines representations that encode lexical content
  and latent user attributes such as identity, intent, background, or ideology.
  These semantic representations are central to the power of LLMs, yet they also
  introduce a unique class of privacy vulnerabilities. Unlike traditional data
  privacy breaches, semantic privacy threats do not rely on direct access to raw
  data or identifiers. Instead, they exploit the model’s ability to retain and regenerate
  meaning, enabling adversaries to infer sensitive information from internal
  representations or observable outputs. This section unpacks how semantic information
  flows across the LLM lifecycle and introduces four representative attack vectors
  - membership inference, attribute inference, model inversion, and backdoor
  attacks - that target distinct stages and components of the semantic
  processing chain, as shown in Fig.~\ref{fig:LLMstructure&Attack}.

  \subsection{Semantic Information in LLM}

  \subsubsection{S1: Input Processing}

  Input processing marks the entry point of semantic transformation within an
  LLM. The raw input text undergoes tokenization, a process that dissects
  natural language into discrete subword units based on statistical segmentation
  techniques like Byte Pair Encoding or WordPiece~\cite{huang2024large}. These tokens
  preserve core semantic constituents through carefully trained vocabulary mappings.
  These tokens are then mapped to dense vectors in a continuous space via
  embedding layers. Positional encoding is subsequently added to preserve syntactic
  order, allowing the model to differentiate between semantically distinct
  permutations. Together, embeddings and positional encodings instantiate the
  semantic structure of the input in a high-dimensional space suitable for
  subsequent reasoning.

  \subsubsection{S2: Pretraining}

  Pretraining occurs over multiple stacked transformer layers, each consisting
  of self-attention, feedforward neural networks (FFNs), residual connections, and
  layer normalization~\cite{wang2023pre}. These components jointly enable the
  model to abstract hierarchical semantics from its input, moving from lexical and
  syntactic cues toward higher-order concepts such as intent, logical entailment,
  causality, and commonsense knowledge. Self-attention mechanisms are particularly
  crucial, as they enable the model to construct context-dependent semantic
  relationships across all token positions. FFNs and normalization layers reinforce
  these abstractions, producing increasingly entangled and generalizable
  semantic representations. At the end of pretraining, the model holds a rich
  and implicit understanding of language semantics, encompassing topics,
  sentiments, and factual knowledge.

  \subsubsection{S3: Fine-tuning}

  Fine-tuning adapts the pretrained model to specific downstream tasks by
  training new or modified output heads (e.g., task classifiers, generative
  decoders, or question-answering heads)~\cite{chang2024survey}. At this stage,
  the general-purpose semantic representations learned during pretraining are
  reoriented toward a specific operational goal. The output head translates latent
  semantic features into actionable predictions, such as labels or token
  distributions. Because fine-tuning typically occurs on smaller, curated
  datasets, the model’s semantic alignment with domain-specific language and logic
  becomes more pronounced.

  \subsubsection{S4: Alignment}

  The final stage of LLM deployment is alignment, wherein the model is tuned to conform
  to ethical, social, and safety norms. This includes structured interventions
  such as reinforcement learning from human feedback, output filtering, and
  inference-time controls~\cite{khamassi2024strong}. During inference, the model
  generates outputs that are semantically grounded in both the input and its
  internal representations. Alignment ensures that outputs adhere to acceptable behaviors
  — avoiding toxicity, bias, or unsafe advice — while retaining the semantic intent
  of user prompts. This step is where semantic representations become publicly
  observable, rendering any residual privacy vulnerabilities fully manifest.

  \subsection{Semantic Privacy Threats}

  This section provides a comparative overview (Table~\ref{tab:semantic_attacks})
  of four prominent semantic privacy attacks, including Membership Inference,
  Attribute Inference, Model Inversion, and Backdoor Attacks.

\begin{table}[!htpb]
    \centering
    \caption{Comparison of Semantic Privacy Attacks}
    \label{tab:semantic_attacks}
    \resizebox{\linewidth}{!}{
    \begin{tabular}{p{1.5cm}|c|c|p{3.9cm}}
      \toprule
      \multicolumn{1}{c}{\makecell{\textbf{Attack} \\ \textbf{type}}} &
      \multicolumn{1}{c}{\textbf{Stage}} &
      \multicolumn{1}{c}{\rotatebox{0}{\textbf{Visibility}}} &
      \multicolumn{1}{c}{\makecell{\textbf{Quantifiable} \\\textbf{Metrics}}} \\
      \hline
      \textbf{\includegraphics[height=1em]{fig/membership_inference.png}~\cite{mozaffari2024semantic,he2025towards}} & S4    & \blackbox  & Precision, perplexity, similarity \\
      \hline
      \textbf{\includegraphics[height=1em]{fig/attribute_inference.png}~\cite{hui2024pleak,zheng2024inputsnatch}}    & S3–S4 & \graybox   & AUC, clustering, accuracy         \\
      \hline
      \textbf{\includegraphics[height=1em]{fig/model_inversion.png}~\cite{zhang2024extracting,qu2025prompt}}        & S2–S4 & \whitebox & BLEU, recovery rate, inversion success \\
      \hline
      \textbf{\includegraphics[height=1em]{fig/backdoor_attack.png}~\cite{zhang2024instruction,liu2025compromising}}& S2–S3 & \whitebox & ASR, stealthiness, accuracy       \\
      \bottomrule
    \end{tabular}
    }
    \begin{tablenotes}
      \item[] \blackbox~: Blackbox; \graybox~: Graybox; \whitebox~: Whitebox.
    \end{tablenotes}
\end{table}

\subsubsection{\includegraphics[height=1em]{fig/membership_inference.png} Membership Inference Attack}

  Membership Inference Attacks (MIAs) aim to determine whether specific inputs, such
  as sentences or documents, were used in model training~\cite{song2024not}. LLMs’
  memorization tendencies lead to distinguishable behaviors for seen versus unseen
  data, reflected in confidence scores or response structure. Recent work enhances
  MIAs by exploiting semantic perturbations~\cite{mozaffari2024semantic},
  stochastic embeddings~\cite{galli2024noisy}, and memorization traits beyond overfitting~\cite{fu2024membership},
  improving attack robustness across model types and datasets.

  LLMs can reproduce verbatim or semantically enriched training content~\cite{ippolito2022preventing,
  huang2024demystifying}, allowing adversaries to exploit semantic memorization.
  Classification models rely on confidence scores, while generative models use metrics
  like perplexity or entropy. He et al.~\cite{he2025towards} propose a label-only
  MIA based on semantic similarity, matching logit-based methods. Wen et al.~\cite{wen2024membership}
  show in-context learning is especially vulnerable, and Song et al.~\cite{song2024not}
  introduce likelihood-based attacks targeting hard-to-predict tokens without
  auxiliary data. These works highlight the increasing subtlety and effectiveness
  of semantic-level MIAs.

\subsubsection{\includegraphics[height=1em]{fig/attribute_inference.png} Attribute
Inference Attack}
Attribute inference attacks target the internal representations of ML models,
particularly those trained on natural language data, to infer latent attributes
of the input data that are not explicitly presented. These attributes could
include age, gender, location, political affiliation, or even psychological
traits. Such attacks allow adversaries to extract personal characteristics from
seemingly anonymized or generic inputs, potentially violating user privacy
without any direct data breach.

For example, zheng et al.~\cite{zheng2024inputsnatch} introduces a timing-based
side-channel attack that exploits cache-sharing mechanisms in LLM inference to
steal private inputs, by employing machine learning techniques for vocabulary correlation
and statistical time fitting. Additionally, attackers can exploit the correlation
between certain linguistic features and demographic attributes. By
constructing input queries that vary specific attributes while keeping the
core semantics constant, the attacker can analyze changes in the model’s outputs
or intermediate representations. For example, training shadow models to mimic the
victim model and perform attribute classification based on outputs or embeddings~\cite{hui2024pleak}.

\subsubsection{\includegraphics[height=1em]{fig/model_inversion.png} Model
Inversion Attack}

  Model inversion attacks allow an adversary to infer sensitive information
  about the data used to train a machine learning model by analyzing its outputs~\cite{aguilera2025llm}.
  By analysing model internal states, such as its embeddings, intermediate representations,
  or the output for a given input, adversaries can observe LLMs' responses to
  different queries and exploit. Then, the attackers can reverse-engineer the
  model to reconstruct certain attributes or features of the training data~\cite{tragoudaras2025information}.
  This type of attack is especially concerning for models trained on personal data,
  such as those used in finance, healthcare, and social media platforms, where users'
  sensitive attributes (e.g., income, health conditions) could be inferred from the
  model’s internal representations.

  For example, in collaborative inference scenarios, the work in~\cite{qu2025prompt}
  demonstrates the ability to recover input prompts through transmitted intermediate
  activations. Additionally, the method in~\cite{zhang2024extracting} extracts prompts
  using only text outputs from normal user queries, showcasing zero-shot
  transferability across different LLMs.

\subsubsection{\includegraphics[height=1em]{fig/backdoor_attack.png} Backdoor
Attack}

  In backdoor attacks, an adversary poisons the training process by injecting
  crafted inputs that include a hidden ``trigger'' — a specific pattern or
  sequence in the input data that causes the model to behave abnormally~\cite{yang2024comprehensive}.
  In most cases, the model performs normally for clean inputs, but when it encounters
  inputs containing the trigger, it produces attacker-specified outputs. As semantic
  privacy focuses on the preservation of meaning and context within the data,
  backdoor attacks can manipulate semantic information by embedding triggers
  that exploit the model's understanding of language and context.

  For example, Zhang et al., introduce a novel method of embedding backdoors in customized
  LLMs through semantic-level instructions~\cite{zhang2024instruction}, which do
  not require modifications to the input sentences, thereby enhancing the stealthiness
  of the attack. Liu et al., launch contextual backdoor attacks that can exploit
  adversarial in-context generation to optimize poisoned demonstrations and
  compromise the contextual environment, resulting in context-dependent defects in
  generated programs~\cite{liu2025compromising}.

  \smallskip
  \begin{center}
    \colorbox{teal!10}{
    \begin{minipage}{0.95\linewidth}
      \textbf{Takeaways.} The semantic processing pipeline of LLMs introduces
      multiple stages where sensitive user information may be transformed, retained,
      and ultimately exposed. We noticed that a spectrum of attacks exploits these
      vulnerabilities: Membership Inference identifies training data presence
      via output behaviors; Attribute Inference extracts latent user traits from
      internal model states; Model Inversion reconstructs input features through
      observable outputs or intermediate activations; and Backdoor Attacks manipulate
      model behavior through stealthily embedded semantic triggers.
    \end{minipage}

    }
  \end{center}

  \section{Semantic Privacy Protection}
  \label{Existing LPPMs in vehicular networks}

  Given the multifaceted threats to semantic privacy in LLMs, defending against
  such attacks requires more than generic anonymization or surface-level noise
  injection. Because semantic information is transformed and retained across distinct
  processing stages, protection mechanisms should be stage-aware, representation-sensitive,
  and context-adaptive. A range of emerging techniques aim to safeguard semantic
  privacy by perturbing inputs, encrypting internal states, localizing computation,
  and erasing sensitive knowledge post hoc. However, these solutions vary
  significantly in their efficacy, overhead, and stage-specific applicability. This
  section provides a systematic review of semantic privacy protection strategies,
  highlighting their mechanisms, deployment stages, and limitations. We group existing
  approaches into five broad categories: DP-based
  perturbations, Edge-centric and split architectures, Encrypted representation
  learning, Knowledge Unlearning, and Privacy-aware model pruning and
  compression (Table~\ref{tab:semantic_privacy_summary_extended}).

\begin{table}[!ht]
  \centering
  \caption{Comparison of Semantic Privacy Techniques}
  \label{tab:semantic_privacy_summary_extended}
  \resizebox{\linewidth}{!}{
  \begin{tabular}{p{3.9cm}|p{1.1cm}|c|p{1.6cm}}
    \toprule
    \multicolumn{1}{c}{\textbf{Technique}} &
    \multicolumn{1}{c}{\makecell{\textbf{Targeted} \\ \textbf{threats}}} &
    \multicolumn{1}{c}{\textbf{Stage}} &
    \multicolumn{1}{c}{\makecell{\textbf{Protection} \\ \textbf{capability$^{1}$}}} \\
    \hline
    DP Embedding~\cite{mai2023split} &
    \includegraphics[height=1em]{fig/membership_inference.png}%
    \includegraphics[height=1em]{fig/attribute_inference.png} &
    S1-2 &
    \includegraphics[height=1em]{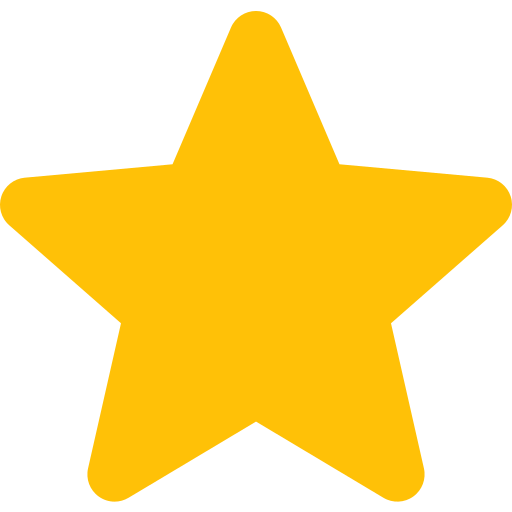}%
    \includegraphics[height=1em]{fig/star.png}%
    \includegraphics[height=1em]{fig/star.png}%
    \includegraphics[height=1em]{fig/star.png} \\
    \hline
    DP Tokenization~\cite{tong2025inferdpt} &
    \includegraphics[height=1em]{fig/membership_inference.png}%
    \includegraphics[height=1em]{fig/attribute_inference.png} &
    S1-2 &
    \includegraphics[height=1em]{fig/star.png}%
    \includegraphics[height=1em]{fig/star.png}%
    \includegraphics[height=1em]{fig/star.png} \\
    \hline
    DP Prompt Encoding~\cite{yu2024textual} &
    \includegraphics[height=1em]{fig/membership_inference.png}%
    \includegraphics[height=1em]{fig/attribute_inference.png}%
    \includegraphics[height=1em]{fig/model_inversion.png} &
    S1,3 &
    \includegraphics[height=1em]{fig/star.png}%
    \includegraphics[height=1em]{fig/star.png}%
    \includegraphics[height=1em]{fig/star.png} \\
    \hline
    Semantic-Aware DP~\cite{wei2023heterogeneous} &
    \includegraphics[height=1em]{fig/attribute_inference.png}%
    \includegraphics[height=1em]{fig/model_inversion.png} &
    S2-3 &
    \includegraphics[height=1em]{fig/star.png}%
    \includegraphics[height=1em]{fig/star.png}%
    \includegraphics[height=1em]{fig/star.png}%
    \includegraphics[height=1em]{fig/star.png} \\
    \hline
    Split Learning + LDP~\cite{yao2024split,mai2023split} &
    \includegraphics[height=1em]{fig/membership_inference.png}%
    \includegraphics[height=1em]{fig/attribute_inference.png} &
    S1-2 &
    \includegraphics[height=1em]{fig/star.png}%
    \includegraphics[height=1em]{fig/star.png}%
    \includegraphics[height=1em]{fig/star.png}%
    \includegraphics[height=1em]{fig/star.png}%
    \includegraphics[height=1em]{fig/star.png} \\
    \hline
    Encrypted Embedding~\cite{mishra2024sentinellms} &
    \includegraphics[height=1em]{fig/attribute_inference.png}%
    \includegraphics[height=1em]{fig/model_inversion.png} &
    S2-3 &
    \includegraphics[height=1em]{fig/star.png}%
    \includegraphics[height=1em]{fig/star.png}%
    \includegraphics[height=1em]{fig/star.png}%
    \includegraphics[height=1em]{fig/star.png} \\
    \hline
    Homomorphic Encryption~\cite{ugurbil2025fission} &
    \includegraphics[height=1em]{fig/membership_inference.png}%
    \includegraphics[height=1em]{fig/model_inversion.png} &
    S2-3 &
    \includegraphics[height=1em]{fig/star.png}%
    \includegraphics[height=1em]{fig/star.png}%
    \includegraphics[height=1em]{fig/star.png}%
    \includegraphics[height=1em]{fig/star.png}%
    \includegraphics[height=1em]{fig/star.png} \\
    \hline
    Semantic Compression~\cite{kale2024texshape} &
    \includegraphics[height=1em]{fig/attribute_inference.png}%
    \includegraphics[height=1em]{fig/model_inversion.png} &
    S2-4 &
    \includegraphics[height=1em]{fig/star.png}%
    \includegraphics[height=1em]{fig/star.png}%
    \includegraphics[height=1em]{fig/star.png} \\
    \hline
    Latent Unlearning~\cite{yu2025unierase} &
    \includegraphics[height=1em]{fig/membership_inference.png}%
    \includegraphics[height=1em]{fig/attribute_inference.png}%
    \includegraphics[height=1em]{fig/backdoor_attack.png} &
    S3-4 &
    \includegraphics[height=1em]{fig/star.png}%
    \includegraphics[height=1em]{fig/star.png}%
    \includegraphics[height=1em]{fig/star.png}%
    \includegraphics[height=1em]{fig/star.png} \\
    \hline
    DP Forward Fine-Tuning~\cite{du2023dp} &
    \includegraphics[height=1em]{fig/membership_inference.png}%
    \includegraphics[height=1em]{fig/attribute_inference.png} &
    S3-4 &
    \includegraphics[height=1em]{fig/star.png}%
    \includegraphics[height=1em]{fig/star.png} \\
    \hline
    Gradient Unlearning~\cite{yao2024large} &
    \includegraphics[height=1em]{fig/membership_inference.png}%
    \includegraphics[height=1em]{fig/backdoor_attack.png} &
    S3-4 &
    \includegraphics[height=1em]{fig/star.png}%
    \includegraphics[height=1em]{fig/star.png}%
    \includegraphics[height=1em]{fig/star.png} \\
    \hline
    Entropy Pruning~\cite{nikitin2024kernel} &
    \includegraphics[height=1em]{fig/backdoor_attack.png}%
    \includegraphics[height=1em]{fig/model_inversion.png} &
    S3-4 &
    \includegraphics[height=1em]{fig/star.png}%
    \includegraphics[height=1em]{fig/star.png} \\
    \hline
    Semantic Distillation~\cite{deusser2025resource} &
    \includegraphics[height=1em]{fig/attribute_inference.png}%
    \includegraphics[height=1em]{fig/model_inversion.png} &
    S2-3 &
    \includegraphics[height=1em]{fig/star.png}%
    \includegraphics[height=1em]{fig/star.png}%
    \includegraphics[height=1em]{fig/star.png} \\
    \hline
    Gradient Masking~\cite{zhang2025exploring} &
    \includegraphics[height=1em]{fig/backdoor_attack.png}%
    \includegraphics[height=1em]{fig/model_inversion.png} &
    S2-3 &
    \includegraphics[height=1em]{fig/star.png}%
    \includegraphics[height=1em]{fig/star.png}%
    \includegraphics[height=1em]{fig/star.png} \\
    \bottomrule
  \end{tabular}
  }
  \begin{tablenotes}
    \small \item[] $^1$ The semantic privacy protection from very low to very high is marked from 1 to 5 \includegraphics[height=1em]{fig/star.png}.
  \end{tablenotes}
\end{table}

  \subsection{Data Locality and Edge Processing}
  Minimizing data transmission by performing computations at the edge significantly
  reduces the exposure of sensitive semantic content to centralized infrastructures.
  This design enhances privacy by keeping both raw inputs and intermediate
  representations within user-controlled environments, thereby mitigating the risks
  of inference attacks and semantic leakage during model execution. Split
  learning is a prominent realization of this paradigm. By partitioning model layers,
  executing early computations (e.g., embeddings or shallow transformers) on the
  client, and transmitting only encoded features to the server, it limits the
  exposure of semantically rich data. Advanced frameworks like Split-N-Denoise~\cite{mai2023split}
  further enhanced privacy by injecting LDP noise on client-side embeddings and applying
  denoising post-inference, thereby forming a semantic bottleneck that weakens
  upstream inference attacks.

  Empirical studies confirm the resilience of split learning to reconstruction
  threats. Yao et al.~\cite{yao2024split} showed that adversarial attacks like
  UnSplit fail to extract meaningful content from intermediate representations
  in transformer-based LLMs, especially when techniques such as dropout and normalization
  are applied in early layers. However, running partial LLMs on client devices imposes
  computational and memory overheads, particularly in resource-constrained
  settings. Moreover, the effectiveness of privacy protection hinges on the
  location of the split point: the shallower a split exposes sensitive semantics,
  while the deeper increases the client-side burden.

  Complementary to architectural partitioning, personalized model updates offer an
  alternative approach to enhancing semantic privacy. By enabling local fine-tuning
  and adaptation without sharing gradients or updates with a centralized model, user-specific
  knowledge remains confined to the edge. For instance, SAP~\cite{shen2023split}
  combined edge-based fine-tuning with selective token-level privatization,
  perturbing only those embeddings deemed utility-relevant and sensitive. While effective
  in balancing performance and privacy, this approach relied on accurate token
  sensitivity identification and incurs device-side resource demands. Additionally,
  it presumed a trusted model vendor and secure edge infrastructure—assumptions
  that may not hold under adversarial deployment environments.

  \subsection{Embedding Encryption and Secure Representation Learning}
  Securing embeddings and intermediate representations is crucial for mitigating
  semantic privacy risks in large language models. Since embeddings encode rich
  semantic content, their exposure can lead to inference attacks, model
  inversion, or the extraction of sensitive training data. Techniques that
  encrypt or obfuscate embeddings aim to prevent adversaries from reconstructing
  inputs or accessing confidential knowledge, thereby offering protection during
  both training and inference phases.

  Homomorphic Encryption (HE), particularly Fully Homomorphic Encryption (FHE), enables
  computation on encrypted embeddings without exposing raw or intermediate data,
  providing strong semantic privacy guarantees. Frameworks like Fission~\cite{ugurbil2025fission}
  adapt a hybrid strategy—utilizing secure multiparty computation for linear operations
  and evaluator networks for non-linear layers, with added data shuffling and
  partitioning to prevent semantic leakage. While theoretically robust, these
  approaches suffered from high computational overheads, limiting their practicality
  in real-time LLM applications and relying on partially trusted infrastructure.

  Complementary strategies focus on minimizing the semantic exposure within embeddings
  themselves. Privacy-preserving embedding compression, as proposed by Kale et al.~\cite{kale2024texshape},
  employed mutual information-guided dimensionality reduction to retain only task-relevant
  features, filtering out sensitive semantics. Though effective in theory, such
  methods depended on accurate priors to distinguish sensitive from non-sensitive
  information and risk performance degradation if compression is too aggressive.
  Alternatively, token and embedding obfuscation methods, such as those in Mishra
  et al.~\cite{mishra2024sentinellms}, leveraged irreversible transformations,
  keyed tokenization, randomized indexing, and spatial embedding distortion, to enable
  encrypted inference without ever decrypting the input. While promising against
  inversion and similarity attacks, this approach assumed a secure and well-aligned
  encryption setup and may face compatibility challenges across diverse LLM architectures.

  \subsection{Differential Privacy}

  DP is fundamental for mitigating privacy risks during
  LLM training by injecting calibrated noise into input data, thus preventing overfitting
  to sensitive linguistic patterns and guarding against attacks like membership inference,
  attribute leakage, and model inversion~\cite{song2024not}. These protections
  extend to semantic privacy by disrupting latent representations that could
  reveal user identity.

  To enhance semantic protection, DP at the embedding level introduces mechanisms
  such as Local Differential Privacy (LDP), randomized transformations, and encrypted
  perturbations. For example, Mai et al. \cite{mai2023split} applied client-side
  LDP with local denoising, while Wang et al. \cite{wang2024selective} perturbed
  only sensitive segments, offering fine-grained control dependent on accurate sensitivity
  detection and noise calibration.

  Beyond embeddings, differentially private tokenization introduces randomness
  into token selection to mitigate memorization and embedding inversion, as shown
  by~\cite{wang2024selective, tong2025inferdpt}, though these methods often
  impact task fidelity. Prompt-level DP has also been explored. Yu et al. \cite{yu2024textual}
  anonymized entities via DP hashes using NER, and Tong et al. \cite{tong2025inferdpt}
  applied perturbed prompt generation. However, both approaches depend heavily on
  accurate entity detection.

  To overcome uniform DP limitations, semantic-aware DP frameworks adapt
  perturbations to semantic structure. Wei et al.~\cite{wei2023heterogeneous}
  proposed a two-stage heterogeneous graph model combining feature attention and
  gradient perturbation with independently budgeted noise. While improving semantic
  alignment, such methods require complex semantic modeling and may degrade
  performance under high noise budgets.

  \subsection{Model Pruning and Compression for Privacy}
  Model pruning and compression techniques play a dual role in enhancing privacy
  and improving computational efficiency. By eliminating redundant parameters,
  these methods reduce the model’s capacity to memorize and retain fine-grained
  training data, thus mitigating semantic privacy risks. Privacy-aware pruning
  methods, such as the mutual information-based strategy by Huang et al.~\cite{huang2024large},
  identified and removed structurally redundant neurons without requiring labeled
  data or retraining, reducing the risk of re-exposing private content. Structured
  pruning approaches like ZipLM~\cite{kurtic2023ziplm} incorporated saliency-based
  criteria to maintain utility while improving latency. However, such methods relied
  on heuristic importance scores and inadvertently removed privacy-benign but task-critical
  components, especially in compact models.

  Embedding compression provides another avenue for controlling semantic
  exposure. By applying low-rank factorization, sparsification, or hashing,
  models can discard detailed lexical and syntactic features while preserving
  essential semantics. Gilbert et al.~\cite{gilbert2023semantic} introduced
  semantic reconstruction effectiveness as a metric to evaluate how well
  compressed embeddings retain meaning, showing that models like GPT-4 can maintain
  utility post-compression. While such methods suppressed the capacity to encode
  identifiable cues, they offered no strong guarantees of recoverability,
  aggressive compression can impair task performance, and residual information may
  still be exploitable by sophisticated inversion attacks.

  Entropy-based pruning further enhances privacy by removing components associated
  with high semantic uncertainty or risk. Techniques like Kernel Language
  Entropy~\cite{nikitin2024kernel} and semantic entropy estimators~\cite{farquhar2024detecting}
  assessed variability across semantically equivalent outputs to guide pruning decisions.
  These methods suppressed the generation of hallucinated content that could
  inadvertently leak sensitive information. However, their effectiveness
  depended on the robustness of semantic similarity metrics, which underperformed
  in complex or domain-diverse language scenarios.

  Finally, compressed secure model distillation and gradient masking offer
  targeted defenses against semantic leakage. Work in~\cite{deusser2025resource}
  distilled large LLMs into encoder-only student models using anonym\-ization-aware
  supervision (e.g., NER-guided masking), preserving high-level semantics while
  removing reconstructive details. Though suitable for deployment in privacy-sensitive
  settings, the pipeline's relianced on annotation quality can result in missed
  entities, particularly in multilingual or informal inputs. Similarly, GradMLMD~\cite{zhang2025exploring}
  introduced gradient-guided masking to suppress exposure of semantically sensitive
  tokens during training. While not explicitly designed for privacy, it reduced overfitting
  to sensitive signals and resists model inversion. Yet, its reliance on accurate
  gradient estimation and lack of formal guarantees limited its robustness in adversarial
  contexts.

  \subsection{Knowledge Unlearning}
  Knowledge unlearning techniques aim to selectively remove the influence of specific
  data points or behaviors from trained language models without retraining from scratch.
  These approaches are increasingly critical for enforcing data privacy rights, such
  as the ``right to be forgotten," and preventing persistent retention of sensitive
  information. Beyond compliance, unlearning contributes to semantic privacy by eliminating
  internal representations and decision pathways tied to the forgotten data,
  ensuring it no longer semantically influences model behavior.

  One class of techniques focuses on privacy-aware fine-tuning. Du et al.~\cite{du2023dp}
  proposed DP-Forward, injecting analytically designed matrix-valued noise into the
  forward pass embeddings to ensure LDP during both training
  and inference. Compared to traditional DP-SGD, this method offered stronger protection
  against semantic leakage via embedding inversion or attribute inference, with
  improved computational efficiency. However, its effectiveness depended on
  precise noise calibration and degraded in tasks with long-range dependencies
  or high-dimensional inputs.

  Other approaches leverage gradient-based unlearning and latent representation
  purging. Yao et al.~\cite{yao2024large} introduced a gradient ascent method
  that reverses the influence of targeted samples, particularly effective in erasing
  harmful or copyrighted content with minimal resources. While suitable for
  semantic unlearning, the method struggled with defining undesirable outputs in
  open-ended generation tasks and lacks guarantees of complete erasure without access
  to original training data. Alternatively, Yu et al.~\cite{yu2025unierase}
  proposed UniErase, which purges semantic traces in a structured latent space using
  vector quantization and sparse autoencoders. By isolating and suppressing
  discrete latent codes linked to sensitive content, the model unlearned specific
  information with minimal utility loss. However, this required accurate identification
  of semantically entangled representations and faced limitations in dense or overlapping
  latent structures.

  \subsection{Alignment and Data Filtering}
  Alignment techniques, particularly those used in instruction tuning and
  reinforcement learning from human feedback (RLHF), play a critical role in shaping
  how LLMs handle sensitive information. As part of the alignment process, training
  data is often curated to exclude harmful, private, or policy-violating content,
  either through automated filtering pipelines or manual annotation. These filtering
  mechanisms aim to prevent the model from learning undesired behaviors or memorizing
  sensitive semantic patterns. However, alignment-stage filtering is inherently limited
  by the quality and granularity of the data selection criteria, overly
  aggressive filters may harm generalization or utility, while insufficiently granular
  ones may allow latent semantic cues to persist. Recent work also explores
  aligning models to privacy-centric objectives, incorporating reward models that
  penalize outputs revealing sensitive attributes or exhibiting identifiable language
  styles. While alignment offers a proactive strategy for semantic privacy
  protection, it requires continuous refinement and auditing to ensure robustness
  against adversarial queries and domain shifts \cite{gong2024privacy, choi2024semantics}.
 \begin{table*}
    [t]
    \centering
    \caption{Challenges and Research Directions for Semantic Privacy in LLMs}
    \label{tab:semantic_privacy_challenges}
    \renewcommand{\arraystretch}{1}
    \begin{tabular}{p{5.5cm}|p{4.5cm}|p{5cm}}
      \toprule \multicolumn{1}{c}{\includegraphics[height=1em]{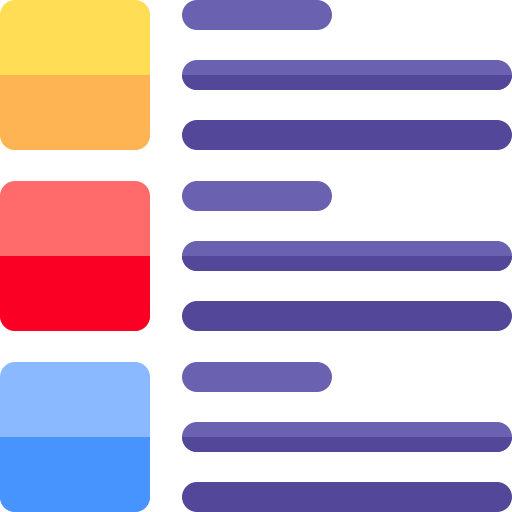} \textbf{Topics}}                    & \multicolumn{1}{c}{\includegraphics[height=1em]{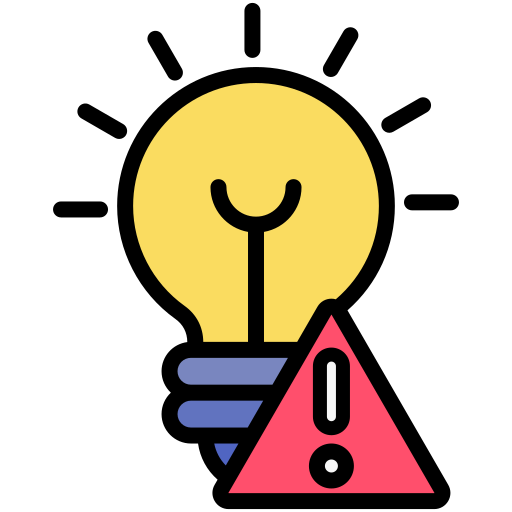} \textbf{Challenges}}      & \multicolumn{1}{c}{\includegraphics[height=1em]{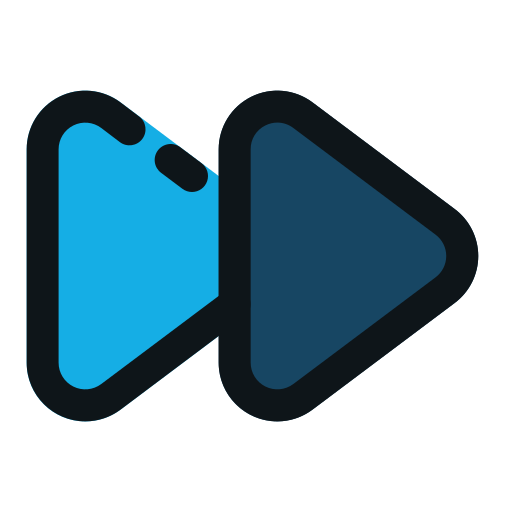} \textbf{Future directions}}                               \\
      \midrule \multirow{3}{*}{\includegraphics[height=1em]{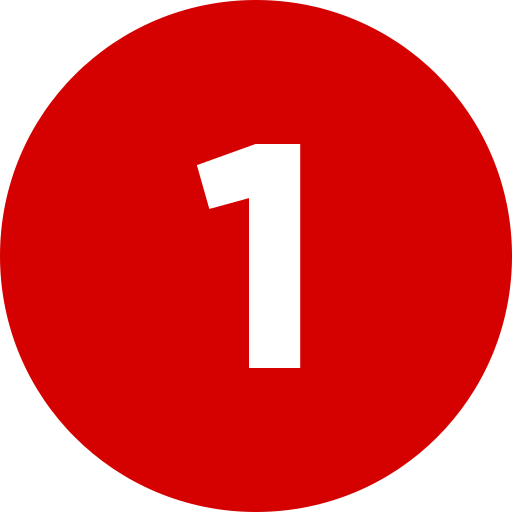} \textbf{Semantic leakage modeling}} & $\bullet$ Implicit leakage                                                                     & $\bullet$ Probabilistic modeling                                                                                        \\
                                                                                                                  & $\bullet$ Contextual reasoning                                                                 & $\bullet$ Inference-time controllers                                                                                    \\
                                                                                                                  & $\bullet$ Stylistic mimicry                                                                    & $\bullet$ Semantic attribution                                                                                          \\
      \hline
      \multirow{3}{*}{\includegraphics[height=1em]{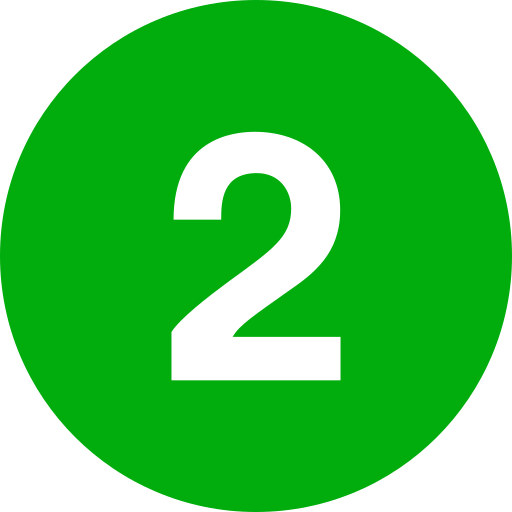} \textbf{Semantic Privacy Quantification}}         & $\bullet$ Cross-sentence semantics                                                             & $\bullet$ Interpretable scoring                                                                                         \\
                                                                                                                  & $\bullet$ Implicit identifiers                                                                 & $\bullet$ Controlled rewriting                                                                                          \\
                                                                                                                  & $\bullet$ Stylistic signals                                                                    & $\bullet$ Leakage risk estimation                                                                                       \\
      \hline
      \makecell[l]{\includegraphics[height=1em]{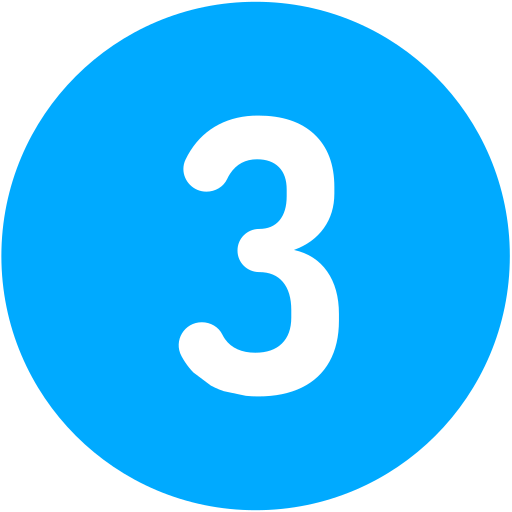} \textbf{Multimodal Privacy}}                       & \makecell[l]{$\bullet$ Cross-modal identity leakage \\ $\bullet$ Image-caption reconstruction} & \makecell[l]{$\bullet$ Multimodal modeling \\ $\bullet$ Adversarial training \\ $\bullet$ Visual-semantic traceability} \\
      \bottomrule
    \end{tabular}
  \end{table*}
  \smallskip
  \begin{center}
    \colorbox{teal!10}{
    \begin{minipage}{0.95\linewidth}
      \textbf{Takeaways.} Protecting semantic privacy in LLMs needs a stage-aware
      approach that combines methods like noise injection, encryption, local
      processing, and forgetting. DP helps protect inputs but
      may reduce meaning. Split learning keeps early steps on local devices for better
      privacy but needs more resources. Techniques like homomorphic encryption
      and semantic compression help prevent data leaks but can be slow or hard
      to integrate. Later-stage methods like unlearning and pruning try to remove
      sensitive features, but their success depends on finding the right ones. No
      single method is enough, effective privacy needs coordination across
      entire model pipeline.
    \end{minipage}

    }
  \end{center}

  \section{Lessons Learned and Open Challenges}
  \label{Location Privacy Challenges in 5G/6G-enabled vehicular networks}

  While recent advances have proposed diverse mechanisms to protect semantic
  privacy in LLMs, significant gaps remain in our understanding and mitigation of
  real-world risks. Existing defenses often focus on syntactic obfuscation or
  noise injection but fail to capture the latent, context-dependent nature of semantic
  leakage, where sensitive information is preserved, inferred, or reconstructed
  without being explicitly reproduced. These limitations are further compounded by
  emerging trends in multimodal modeling, personalized generation, and regulatory
  demand for explainable AI. In this section, we distill key lessons from current
  research and highlight open challenges that should be addressed to advance robust,
  trustworthy semantic privacy, as shown in Table~\ref{tab:semantic_privacy_challenges}.

  \subsection{Modeling of Semantic Information Leakage}

  Existing privacy-preserving mechanisms like DP and cryptographic
  obfuscation primarily target structured data and numerical noise, but fall
  short in addressing semantic leakage in LLMs, where outputs may inadvertently reveal
  sensitive information through stylistic cues, contextual relevance, or abstracted
  summaries. Unlike explicit data leakage, semantic risks arise from the model's
  latent ability to infer or paraphrase private content, often escaping standard
  metrics such as perplexity or token overlap. To address these challenges,
  emerging approaches propose embedding-based similarity analysis, inference-driven
  re-identification, and entailment-aware leakage metrics. Future research should
  explore probabilistic models under adversarial settings and develop inference-time
  privacy controllers, enabling dynamic detection and mitigation of semantic
  risks. Techniques like contrastive language modeling, transformer probing, and
  semantic attribution can further isolate leakage pathways and support audits,
  including retroactive detection of privacy violations even under obfuscation~\cite{yu2025unierase,
  he2025towards}.

  \subsection{Quantification of Semantic Privacy Protection}

  Semantic de-identification aims to obscure sensitive information while preserving
  utility, yet existing methods like named entity anonymization often degrade
  coherence and fail to account for latent, context-dependent identifiers such as
  profession or style. This challenge is amplified in multimodal LLMs, where private
  signals may emerge through cross-modal interactions (e.g., identity leakage
  from visual captions~\cite{gilbert2023semantic}). To address this, emerging techniques
  like privacy-preserving generation, controlled rewriting, and attention-based suppression
  seek to balance privacy with generation quality. Robust evaluation metrics are
  also needed to assess residual identifiability and utility beyond token-level
  measures.

  Equally important is the interpretability and ethical framing of semantic
  privacy mechanisms. Transparent, explainable interventions—such as semantic
  attribution or modality-aware privacy scoring—are essential for accountability
  and trust, especially in sensitive domains. Personalized de-identification
  systems that adapt to user preferences and risk profiles can further support
  ethical deployment. Ultimately, semantic privacy protection should align
  technical effectiveness with social and normative responsibility~\cite{yu2025unierase}.

  \subsection{Multimodal Semantic Privacy Protection}
  Semantic privacy risks in LLMs arise from their capacity to infer and
  regenerate sensitive information via latent associations and contextual
  reasoning, making traditional token-level privacy techniques inadequate.
  Unlike explicit identifiers, semantic cues are often implicit and distributed,
  especially in multimodal settings, complicating detection and modeling. To
  address this, robust semantic privacy quantification should adopt metrics that
  capture meaning preservation, inference likelihood, and contextual identifiability,
  such as embedding similarity, entailment-aware scoring, and re-identification
  risk estimation. At the same time, semantic de-identification should balance
  privacy with generation quality, using methods like controlled rewriting or adversarial
  training to preserve coherence and utility. Transparent, explainable mechanisms
  are also critical to ensure accountability and user trust.

  \smallskip

  \begin{center}
    \colorbox{teal!10}{
    \begin{minipage}{0.95\linewidth}
      \textbf{Takeaways.} Current privacy methods miss how LLMs leak sensitive
      info through reasoning and context. We need better ways to measure and control
      semantic leakage, especially in multimodal models where info spreads
      across text and images.
    \end{minipage}

    }
  \end{center}

  \section{Conclusion}
  \label{conclusion}

  We examine semantic privacy in LLMs, which focuses on the protection of implicit
  and contextually inferred information beyond data privacy in all existing studies.
  We review key attack vectors, including membership inference, attribute
  inference, model inversion, and backdoor attacks across every stage of the LLM
  lifecycle. We discuss the limitations of current defences covering differential
  privacy, embedding encryption, and knowledge unlearning, especially in
  handling contextual inference and semantic ambiguity. We also outline future directions,
  including but not limited to quantifying semantic leakage, multimodal protection,
  privacy-utility trade-offs, and transparent privacy mechanisms.

  \bibliographystyle{ACM-Reference-Format}
  \bibliography{bib}
\end{document}